\documentclass{elsart}
\usepackage{epsfig}
\usepackage{here}
\journal{Astroparticle Physics}

\newcommand{\be}{\begin{equation}}
\newcommand{\ee}{\end{equation}}

\usepackage{amssymb}

\begin{document}

\begin {frontmatter}

\title{On-line recognition of supernova neutrino bursts in the LVD detector}


\author[inr]{N.Yu.Agafonova},
\author[to]{M.Aglietta},
\author[bo]{P.Antonioli},
\author[bo]{G.Bari},
\author[to]{A.Bonardi},
\author[inr]{V.V.Boyarkin},
\author[lngs]{G.Bruno},
\author[to]{W.Fulgione\corauthref{cor}},
\corauth[cor]{Corresponding author: Walter Fulgione, c/o Laboratori Nazionali del Gran Sasso 
S.S. 17 BIS km. 18.910, 67010 Assergi LÕAquila - Italy, email: fulgione@to.infn.it},
\author[to]{P.Galeotti},
\author[bo]{M.Garbini},
\author[lngs]{P.L.Ghia $^{\rm b,}$}, 
\author[bo]{P.Giusti},
\author[to]{ F.Gomez},
\author[br]{E.Kemp},
\author[inr]{V.V.Kuznetsov},
\author[inr]{V.A.Kuznetsov},
\author[inr]{A.S.Malguin},
\author[bo]{H.Menghetti},
\author[bo]{A.Pesci},
\author[bo]{R.Persiani},
\author[mit]{I.A.Pless},
\author[to]{A.Porta},
\author[inr]{V.G.Ryasny},
\author[inr]{O.G.Ryazhskaya},
\author[to]{O.Saavedra},
\author[bo]{G.Sartorelli},
\author[bo]{M.Selvi}
\author[to]{C.Vigorito},
\author[lnf]{L.Votano},
\author[inr]{V.F.Yakushev},
\author[inr]{G.T.Zatsepin},
\author[bo]{A.Zichichi}

\address[inr]{Institute for Nuclear Research, Russian Academy of Sciences, Moscow, Russia}
\address[to]{Institute of Physics of Interplanetary Space, INAF, Torino, University of Torino and INFN-Torino, Italy}
\address[bo]{University of Bologna and INFN-Bologna, Italy}
\address[br]{University of Campinas, Campinas, Brazil}
\address[mit]{Massachusetts Institute of Technology, Cambridge, USA}
\address[lngs]{INFN-LNGS, Assergi, Italy}
\address[lnf]{INFN-LNF, Frascati, Italy}

\begin{abstract}
In this paper we show the capabilities of the Large Volume Detector (INFN Gran Sasso National Laboratory) to identify a neutrino burst associated to a supernova explosion, in the absence of an "external trigger", e.g., an optical observation. We describe how the detector trigger and event selection have been optimized for this purpose, and we detail the algorithm used for the on-line burst recognition. The on-line sensitivity of the detector is defined and discussed in terms of supernova distance and $\bar\nu_e$ intensity at the source.
\end{abstract}


\end{frontmatter}

\normalsize 

\section{Introduction}
The detection of neutrinos from SN1987A marked the beginning of a new phase of neutrino astrophysics \cite{SN1987A-1,SN1987A-2,SN1987A-3,SN1987A-4}.
In spite of the lack of a firmly established theory of core collapse supernova explosion, the correlated neutrino emission appears to be well established.
However, since this first $\nu$ observation was guided by the optical one, the detector capabilities of identifying a $\nu$ burst in the absence of an "external trigger" should be demonstrated very carefully. 
In the presence of an electromagnetic counterpart, on the other hand, the prompt identification of the neutrino signal could alert the worldwide network of observatories allowing study of all aspects of the rare event from its onset.\\ 
The Large Volume Detector (LVD), in the INFN Gran Sasso National Laboratory (Italy), at the depth of 3600 m w.e., is a 1 kt liquid scintillator detector whose major purpose is monitoring the Galaxy to study neutrino bursts from gravitational stellar collapses \cite{LVD}. Besides interactions with protons and carbon nuclei in the liquid scintillator, LVD is also sensitive to interactions with the iron nuclei of the support structure whose total mass is 0.9 kt \cite{oscill}.
The experiment has been taking data, under different configurations, since 1992, reaching in 2001 its present and final configuration.
Its modularity and rock overburden, together with the trigger strategy, make this detector particularly suited to disentangle on-line a cluster of neutrino signals from the background. \\
We will discuss in this paper the LVD performances from the point of view of the on-line identification of a neutrino burst: we will describe in section 2 the trigger of the detector, the event selection and the on-line burst recognition. In section 3 we will define and discuss the detector sensitivity to neutrino bursts which, as we will show in section 4, can be expressed in terms of supernova distance or neutrino intensity at the source.

\section {The LVD event selection chain}
\subsection{The trigger}
\begin{figure}[H]
  \begin{minipage}{.48\columnwidth}
    \begin{center}
      \vspace{-0.cm}
      \includegraphics*[width=7.5cm,angle=0,clip]{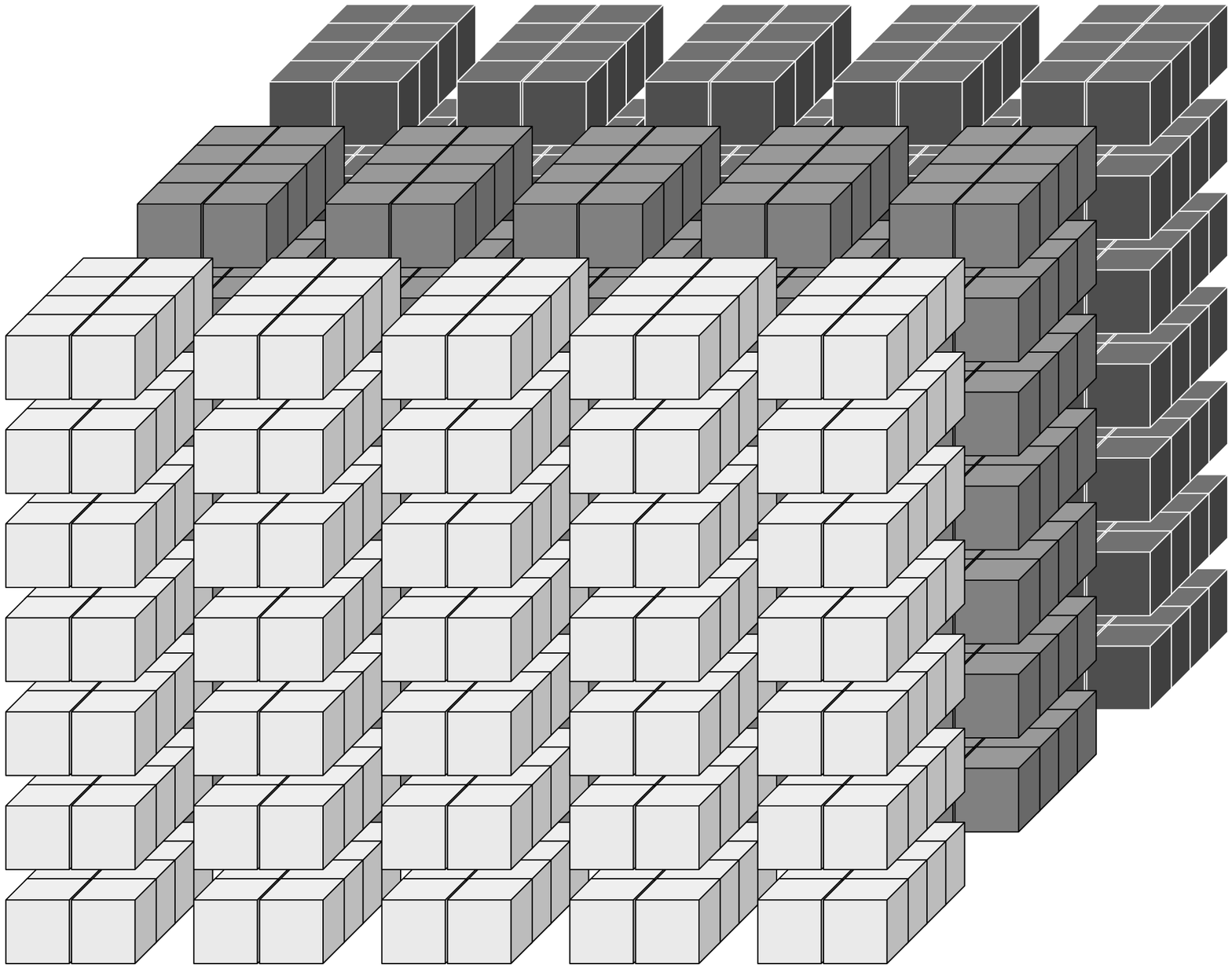}
     \end{center}
   \vspace{0cm}
   \caption{{\bf LVD schematic view: the building block is a module of 8 scintillator counters, 35 modules form a tower.}}
    \label{scatolini}
  \end{minipage}
  \hspace{1pc} 
  \begin{minipage}{.48\columnwidth}
    \begin{center}
       \vspace{0cm}
      \includegraphics*[width=7.5cm,angle=0,clip]{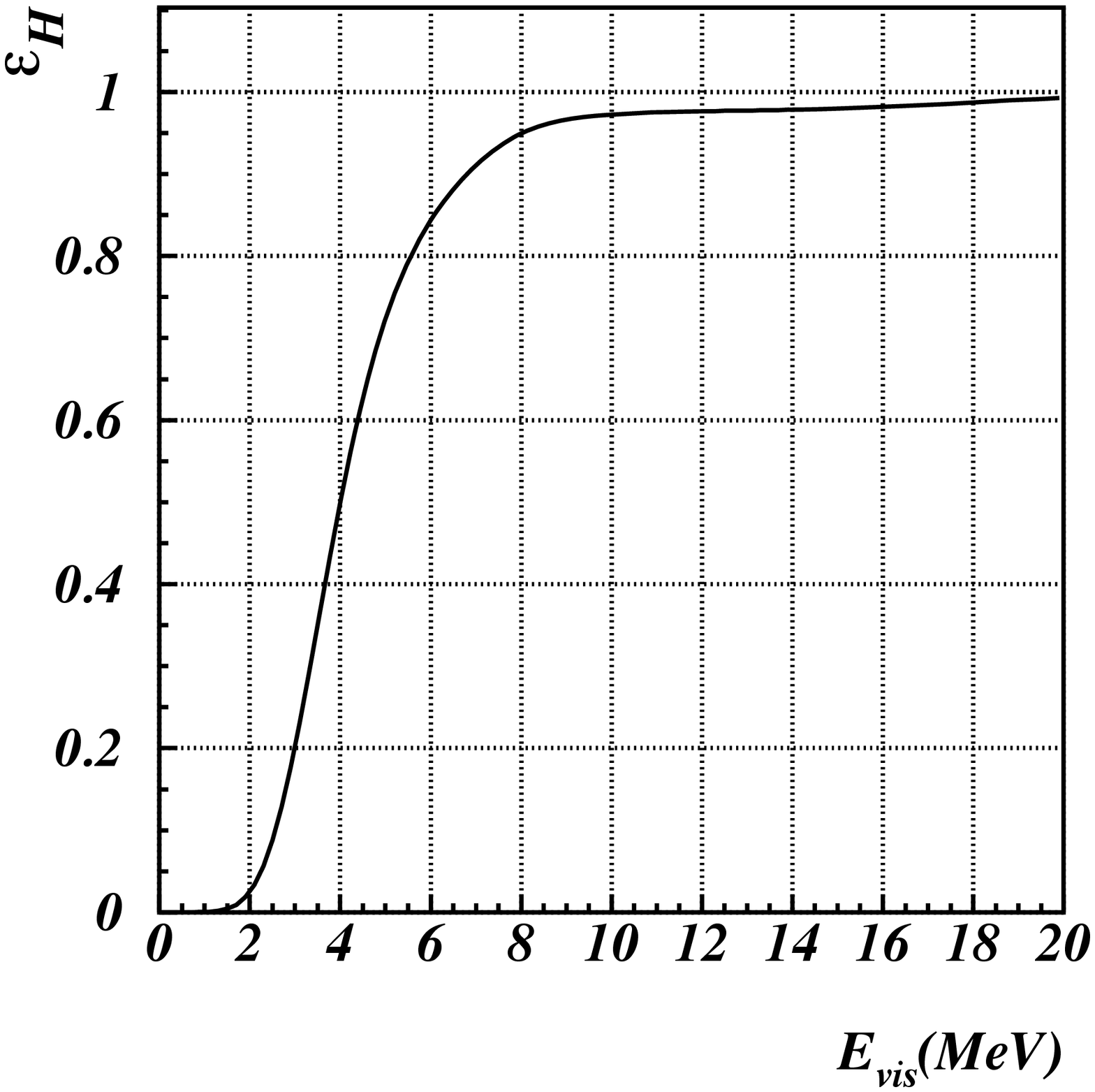}
    \end{center}
    \vspace{-0.cm}
    \caption{{\bf Trigger efficiency, averaged over all the scintillator counters, for the H coincidence versus visible energy $E_\mathrm{vis}$.}}
    \label{threshold1}
  \end{minipage}
\end{figure}

LVD consists of an array of 840 scintillator counters, 1.5  m$^3$ each. The whole array is divided in three identical "towers" with independent high voltage power supply, trigger and data acquisition (see figure \ref {scatolini}).
In turn, each tower consists of 35 "modules" hosting a cluster of 8 counters. 
Each counter is viewed from the top by three 15 cm photomultiplier tubes (PMTs) FEU49b or FEU125.
The charge of the summed PMT signals is digitized by a 12 bit ADC (conversion time = 1 $\mu$s) and the arrival time is measured with a relative accuracy of 12.5 ns and an absolute one of 100 ns \cite{Bigongiari}.\\
The main neutrino reaction in LVD is $\bar \nu_\mathrm{e} \mathrm{p \rightarrow e^+ n}$, which gives two
detectable signals: the prompt one due to the $\mathrm{e}^+$ 
(visible energy $E_\mathrm{vis} \simeq E_{\bar\nu_\mathrm{e}}-1.8$ MeV $+~ 2 ~\mathrm{m_e} 
\mathrm{c}^2$) followed by the signal from the $\mathrm{n p \rightarrow d} \gamma$ capture ($E_{\gamma} = 2.2$ MeV, mean capture time $\simeq 185~\mu \mathrm{s}$,).
The trigger logic is optimized for the detection of both products of the inverse beta decay and is based on the three-fold coincidence of the PMTs of a single counter.
Each PMT is discriminated at two different thresholds resulting in two possible levels of coincidence between a counter's PMTs: H and L, corresponding to ${\cal E}_H \simeq 4$ MeV and ${\cal E}_L \simeq 1$ MeV.
The H coincidence in any counter represents the trigger condition for the array. Once a trigger has been identified, the charge and time of the three summed PMTs' signals are stored in a memory buffer. All signals satisfying the L coincidences in the same module of the trigger counter are also stored, if they occur within 1 ms.
The average efficiency of the H trigger (for electrons) is shown in fig. \ref{threshold1} as a function of the visible energy $E_\mathrm{vis}$. 
The average neutron detection efficiency, $\epsilon_\mathrm{n}$, amounts to about 50\% for neutrons detected in the same counter where the positron has been detected \cite{Amanda}.

One millisecond after the occurrence of a trigger, the memory buffers\footnote{Each memory buffer, one per module, is able to store up to 512 signals, corresponding to 50000 signals in the whole apparatus.}, containing the charge and time information of both H and L signals, are read out.
This is performed independently on the three towers, without introducing any dead time.
The tower event fragments, if present, are then sent to a central processor which provides event-building requiring time coincidence of the first pulse registered in each tower within 10 microsecond. 
The whole duration of one event can extend up to 1 ms.

\subsection{Event selection criteria}
\begin{figure}[H]
  \begin{minipage}{.48\columnwidth}
    \begin{center}
     \vspace{-1.cm}
      \includegraphics*[width=7.5cm,angle=0,clip]{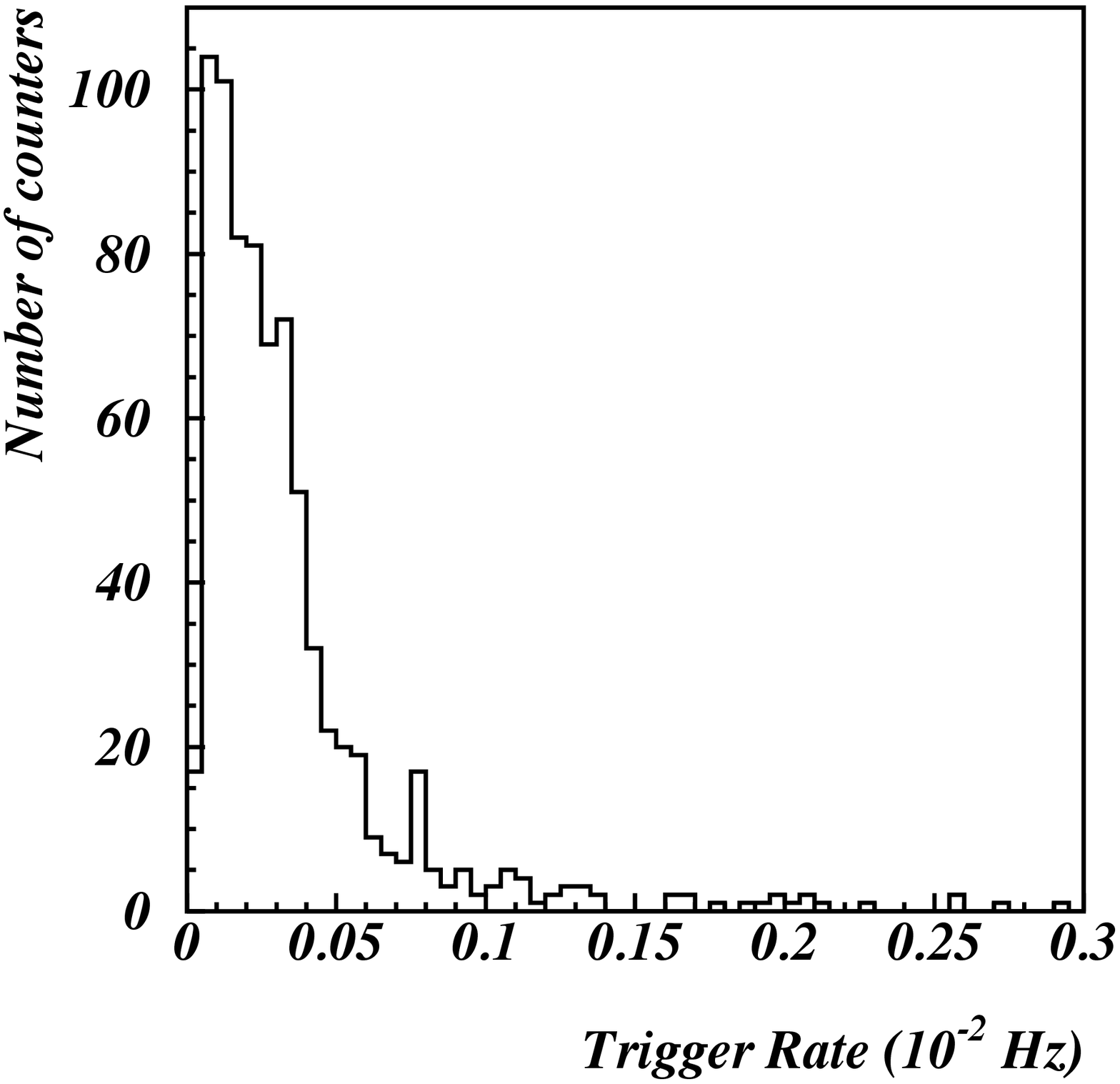}
     \end{center}
     \vspace{-0.cm}
    \caption{{\bf Counting rate distribution (for E$_\mathrm{vis}$ $\geq$ 7 MeV) for all good counters (after cuts), averaged over 150 days of data taking.
    The average rate is $\bar R = 2 \cdot 10^{-4}$ s$^{-1}$ counter$^{-1}$.}}
  \label{rate}  
  \end{minipage}
  \hspace{1pc} 
  \begin{minipage}{.48\columnwidth}
    \begin{center}
     \vspace{-1.cm}
       \includegraphics*[width=7.5cm,angle=0,clip]{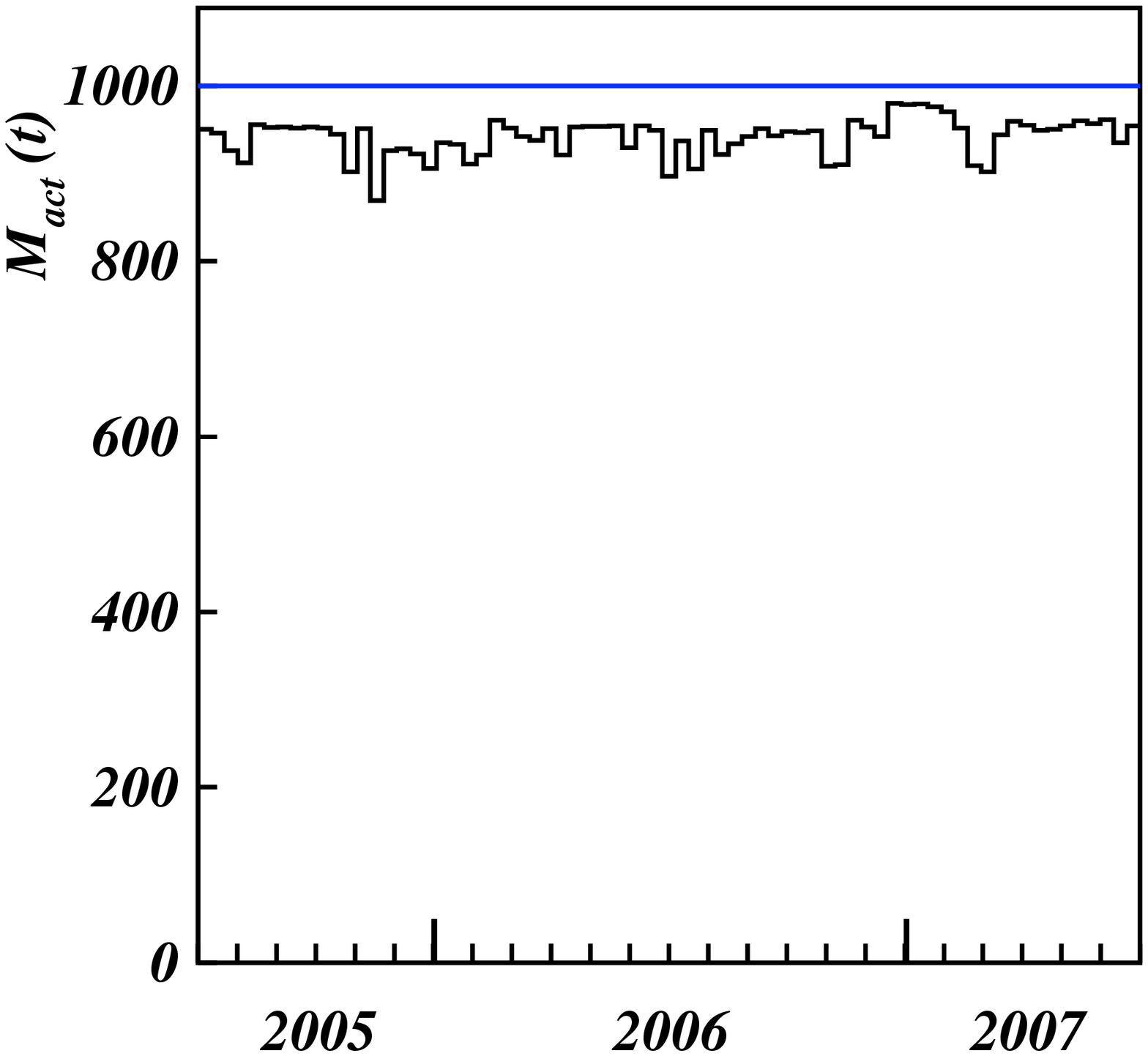}
    \end{center}
    \vspace{-0.cm}
    \caption{{\bf On-line LVD active mass during the last two year of operation.}}
    \label{mass}
  \end{minipage}
\end{figure}

\begin{figure}[H]
  \begin{minipage}{.48\columnwidth}
    \begin{center}
     \vspace{-1.cm}
      \includegraphics[height=7.5cm]{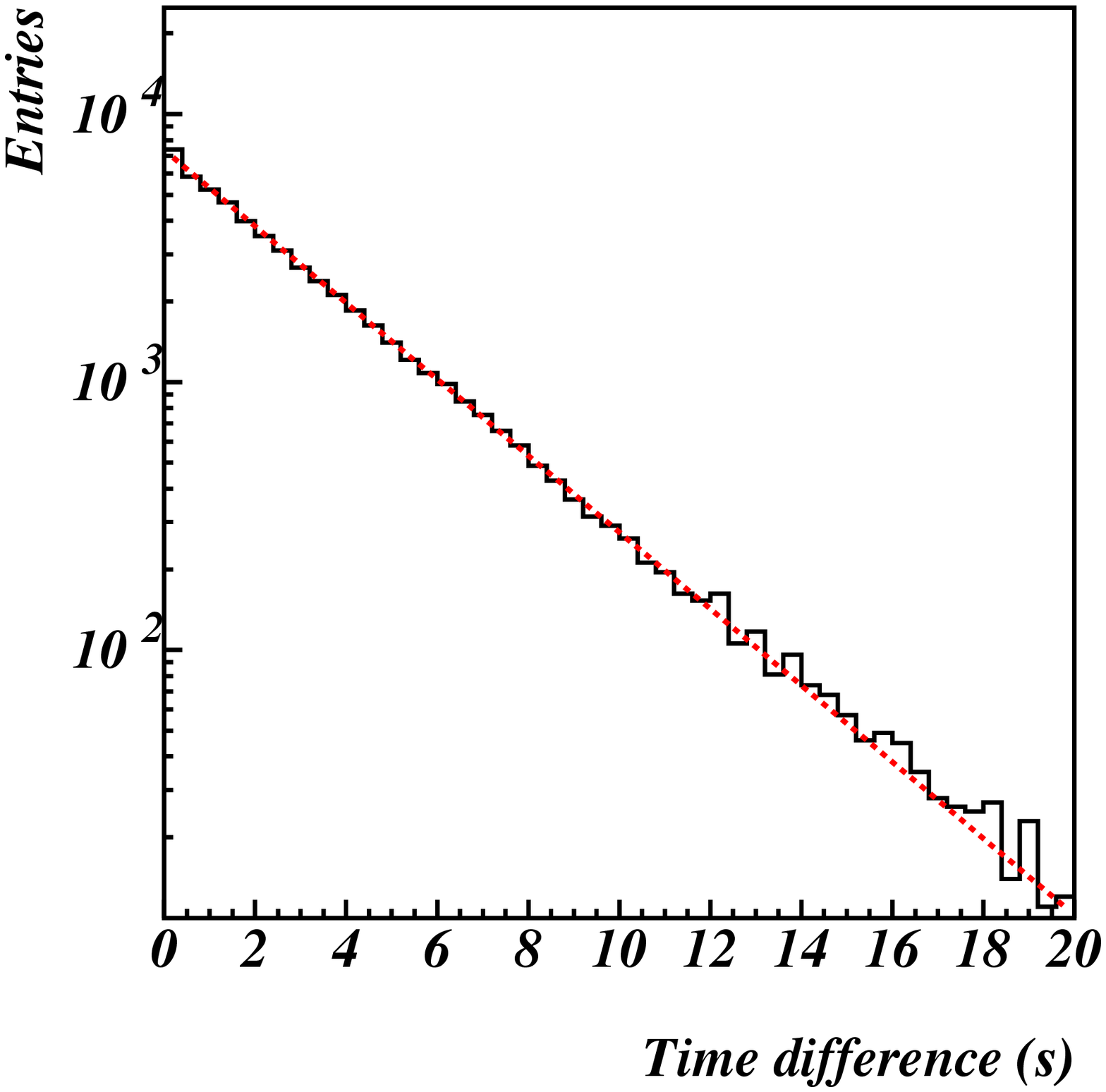}
    \end{center}
    \vspace{0.cm}
   \caption{{\bf Distribution of the difference of the arrival time between successive signals, compared with Poisson expectations.}}
    \label{bdel}
  \end{minipage}
  \hspace{1pc} 
  \begin{minipage}{.48\columnwidth}
    \begin{center}
      \includegraphics[height=7.5cm]{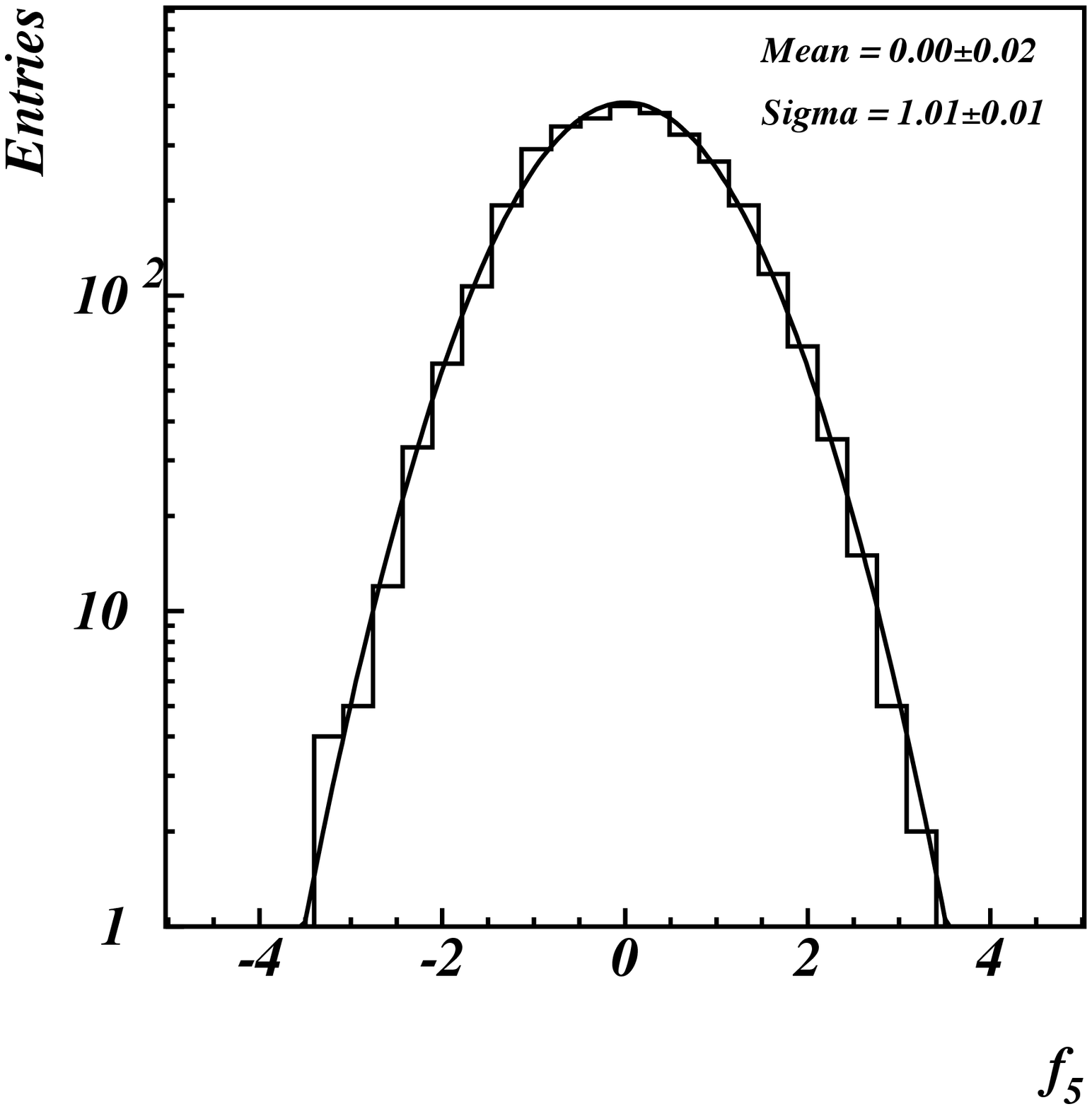}
    \end{center}
    \vspace{0.cm}
    \caption{{\bf Distribution of the fluctuations of the 5 minutes counting rate ( f$_\mathrm{5}= (s_\mathrm{5}-\bar s_\mathrm{5})/\sigma_\mathrm{exp}$). The superimposed curve is the free parameters Gaussian fit. The statistics represents 100 days of data taking.}}
    \label{fluc}
  \end{minipage}
\end{figure}

The basis of the search for neutrino bursts is the identification of clusters of signals in fixed time windows. 
A preliminary step consists in the selection of good signals detected by good counters;
we thus apply to the events two different series of cuts:
(A) to the signals, 
(B) to the counters,
namely:  
\begin{description}
\item[(A) Cut on signals]
\end{description}
\begin{enumerate}
\item The energy of the signals must be in the range E$_\mathrm{cut} \leq$ E$_\mathrm{signal} \leq $100 MeV. Two energy thresholds are considered: E$_\mathrm{cut} = 7$ MeV, corresponds to an average trigger efficiency $> 90\%$, and
E$_\mathrm{cut} = 10$ MeV, to an average trigger efficiency $> 95\%$
(see fig. \ref{threshold1}).
\item Events with signals in coincidence in 2 or more counters of the array, within 200 ns, are rejected
since they are considered muon candidates\footnote{This rejection corresponds to a dead time $\simeq 0.01~\%$, 
taking into account that the global muon rate in LVD is about 0.1 muon s$^{-1}$.}. The probability for electron or positron with energy up to 60 MeV to trigger 2 LVD counters is less than 10\% \cite{Piter}. 
\end{enumerate}
\begin{description}
\item[(B) Cut on counters] 
\end{description}
\begin{enumerate}
\item The response to muons of atmospheric origin is used to identify and discard not properly working counters. The measurement of muons associated with the CNGS beam \cite{CNGS} provides complementary tests of the detection capabilities of the detector.
A wrong muon counting rate or spectrum can be due to scintillator, PMT or electronics problems (ADC or TDC). Counters rejected for this reason are less than 5$\%$ of the total and represents a steady loss of active mass requiring a maintenance intervention. 
\item Counters with a background rate (for E $\geq$ 7 MeV) R $\geq3\cdot 10^{-3}$ s$^{-1}$, during the last two hours of operation, are rejected as noisy. 
The problem arises from electronics or bad energy calibration and usually regards less than $2\%$ of the counters. 
Also in this case counters involved are almost always the same, a new energy calibration or a maintenance intervention are required.
The typical counting rate distribution for all good counters (after cuts), averaged over 150 days of data taking, is shown in figure \ref{rate}.  
\item Counters that take part too often in a cluster of signals are rejected, and the cluster redefined. Namely, if $m$ is the cluster multiplicity and N$_c$ the number of active counters, a counter will be rejected if its multiplicity $m_\mathrm{i}$ corresponds to a probability 
$\sum_\mathrm{k=m_{\mathrm{i}}}^{\infty}$ P(k;m/N$_c$)$ \leq 1 \cdot 10^{-7}$.
\footnote{For example: in a cluster (m=100, $\Delta$t=20 s) detected when there are 800 active counters, a counter will be rejected (and the cluster redefined) if 5 or more signals inside the cluster come from it. } 
This is due to sporadical (of the order of once per month) and local electric noise and determines a mass loss $\ll 0.5\%$.
\end{enumerate}
The effect of the cuts on counters is to adjust dynamically the LVD active mass, M$_\mathrm{act}$, by rejecting counters that are steadily out of order (1 and 2) and/or that are sporadically noisy (3).
A snapshot of the detector active mass in the last two years is shown in figure \ref{mass}.
In the same period the average background counting rate of the whole array was f$_\mathrm{bk}$=0.2 Hz if E$_\mathrm{cut}$=7 MeV and f$_\mathrm{bk}$=0.03 Hz for E$_\mathrm{cut}$=10 MeV.\\
At the end of the whole selection procedure the time distribution of the signals is well described by the Poisson statistics, as can be seen in figure \ref{bdel}, where we show the difference of the arrival time between successive signals, and in figure \ref{fluc} where we show the fluctuations of the 5 min counting rate, s$_\mathrm{5}$, with respect to the local average value (measured during a time window of 40 min) in units of the expected error, $\sigma_\mathrm{exp}$, calculated assuming pure Poisson fluctuations. 
The typical experimental distribution, obtained during 100 days of operation, is fitted with a Gaussian one with mean equal zero and $\sigma$ = 1.01, showing that the residual non-Poissonian contribution to the fluctuations $\sigma_\mathrm{res}=\sqrt{\sigma^2-1}$ is less than 15\%.\\
\subsection{The $\nu$ bursts selection algorithm}
\begin{figure}[H]
  \begin{minipage}{.48\columnwidth}
     \vspace{-0cm}
     \begin{center}
     \includegraphics*[width=0.7\textwidth,angle=90,clip]{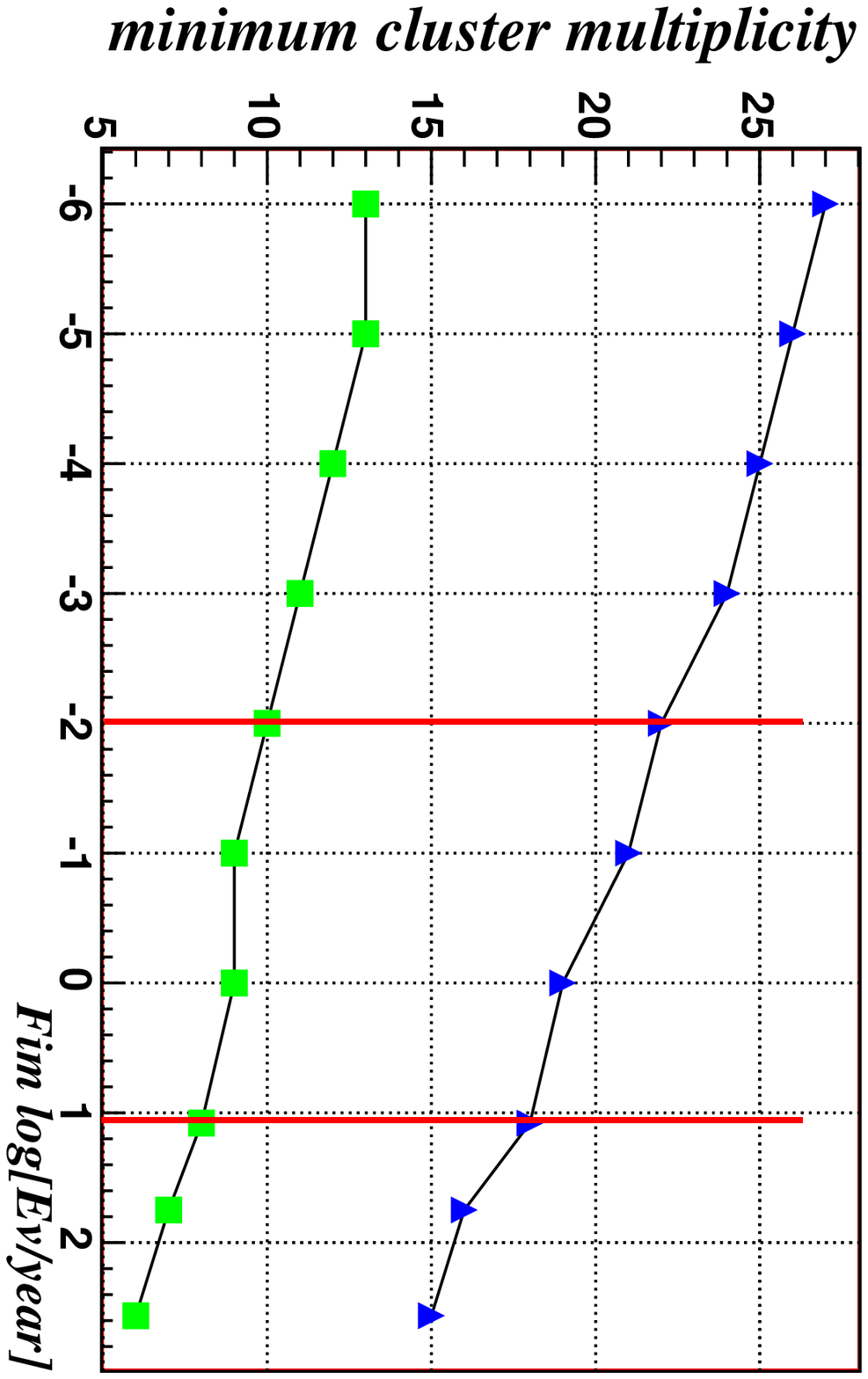}
    \end{center}
    \vspace{0cm}
   \caption{{\bf Minimum cluster multiplicity $m_\mathrm{min}$ vs. the imitation frequency $F_\mathrm{im}$. Triangles correspond to E$_\mathrm{cut}$=7 MeV and f$_\mathrm{bk}$=0.2 Hz, while squares to E$_\mathrm{cut}$=10 MeV and f$_\mathrm{bk}$=0.03 Hz. The two vertical lines represent the $F_\mathrm{im}$ thresholds of 1 candidate per 100 year (stand-alone) and of 1 per month (SNEWS).}}
     \label{mult} 
  \end{minipage}
  \hspace{1pc} 
  \begin{minipage}{.48\columnwidth}
 \vspace{-0cm}
    \begin{center}
     \includegraphics*[width=1.0\textwidth,angle=0,clip]{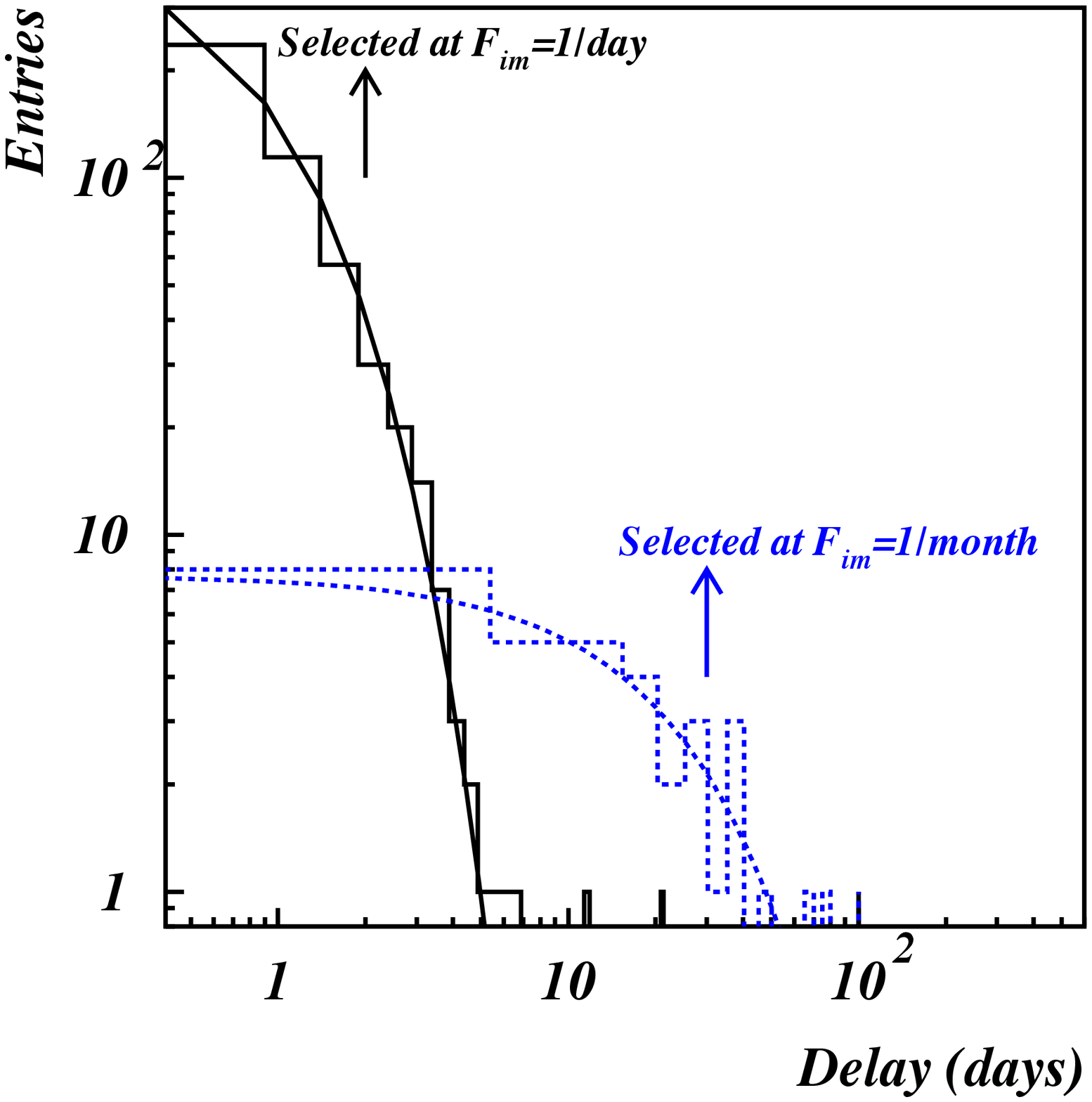}
   \caption{{\bf Distribution of the time intervals between observed clusters (histograms) fitted by Poisson laws (lines) for $F_\mathrm{im}=1$/day (solid) and 
$F_\mathrm{im}=1$/month (dashed), during 688 days and E$_\mathrm{cut}=7~MeV$.}} 
    \label{real} 
    \end{center}
  \end{minipage}
\end{figure}

The core of the algorithm for the on-line selection of candidate neutrino bursts is the search for a cluster of H signals within a fixed-duration time window, $\Delta t$.
The candidate burst is simply characterized by its multiplicity $m$, i.e., the number of pulses detected in $\Delta $t, and by $\Delta $t itself.
All the other characteristics of the cluster, e.g., detailed time structure, energy spectra, $\nu$ flavor content and topological distribution of signals inside the detector are left to a subsequent independent analysis. 
Based on this principle, the LVD data are continuously analyzed by an on-line "supernova monitor".
In detail, each data period, $T$, is scanned through a ``sliding window'' with duration $\Delta $t = 20 s, that is, it is divided into N = 2 $\cdot \frac{T}{\Delta t}-1$  intervals, each one starting in the middle of the previous one, so that the unbiased time window is 10 s. 
The frequency of clusters of duration 20 s and multiplicity $\geq$ m, due to background, is:
\begin{equation}
F_{im}(m,f_\mathrm{bk},20~s) = 8640 \cdot \sum_\mathrm{k\geq m}^{\infty} P(k; 20 \cdot \frac{f_\mathrm{bk}}{s^{-1}}) \ event \cdot day^{-1}
\end{equation}
where f$_\mathrm{bk}$ is the background counting rate of the detector for E$_\mathrm{vis}\geq$ E$_\mathrm{cut}$,
$P(k; f_\mathrm{bk} \Delta t)$ is the Poisson probability to have clusters of multiplicity k if $f_\mathrm{bk} \Delta t$ is the average background multiplicity, and 8640 is the number of trials per day.\footnote{
Once a candidate cluster (m, 20s) has been identified,
the algorithm will search for the most probable starting point of the signal within the time
window. For all possible sub-clusters with multiplicity
$2 \leq k \leq m$ and duration $0 \leq \Delta t \leq 20s$ the
corresponding Poisson probability $P_{k \geq m}$ is calculated and the
absolute minimum identified. The time of the first event of the least probable sub-cluster
is assumed as the start time of the signal.}
For example, a cluster with $m$=10 can be produced by background fluctuations once every 100 years if f$_\mathrm{bk} =$ 0.03 s$^{-1}$; to have the same significance with a higher background, e.g. f$_\mathrm{bk} =$ 0.2 s$^{-1}$, a cluster with much higher multiplicity, $m$=22, would be required (see figure \ref{mult} where $m$ versus $F_\mathrm{im}$ is shown, for two different background conditions). 
For these background rates, $m$ = 10 and $m$ = 22 correspond to the minimum multiplicity, $m_\mathrm{min}$, to have an imitation frequency $F_\mathrm{im} < 1 \cdot 10^{-2}$ yr$^{-1}$.\\
In LVD the search for burst candidates is performed for both energy cuts: 7 and 10 MeV. The chosen $F_{im}$, below which the detected cluster will be an on-line candidate supernova event, is 1 per 100 year working stand-alone while it is relaxed to 1 per month working in coincidence with other detectors, as in the SNEWS project \cite{SNEWS}.\\
In figure \ref{real} we show the distributions of the observed time intervals between selected clusters at the imitation frequency  
$F_\mathrm{im}=1$/day and $F_\mathrm{im}=1$/month during 688 days (between July 5$^\mathrm{th}$, 2005 and May 23$^\mathrm{rd}$, 2007) and E$_\mathrm{cut}=7$ MeV.  
The mean rates, derived from Poisson fits, are $F^{obs}_{im}=1.24$~day$^{-1}$ (with $m_\mathrm{min}$ varying between 13 and 15) and $F^{obs}_{im}=1.28$~month$^{-1}$ ($m_\mathrm{min}$ between 15 and 18), respectively, meaning that, within 25\%,
the time behavior of the background is consistent with expectations.\\
We can conclude that the background trend is predictable even
during long periods of data acquisition and with variable
detector conditions, allowing us to define the significance of a neutrino burst in terms of imitation frequency, $F_{\mathrm{im}}$. \\

\section{Sensitivity to supernova neutrino bursts}
In this section we discuss the sensitivity of the described on-line selection algorithm to the recognition of a supernova event.
The selection method defines a candidate as any cluster of $ m \geq m_\mathrm{min}$ signals within a window of $\Delta t$ = 20 s. For a known background rate, $m_\mathrm{min}$ corresponds to a chosen $F_\mathrm{im}$ which is set as a threshold. This multiplicity represents the minimum number of neutrino interactions required to produce a supernova "alarm", and contains two terms, one due to the background, $f_\mathrm{bk} \Delta t$, and the other due to the neutrino signal.
In particular, for LVD, considering only inverse beta decay (IBD) reactions, which are the dominant ones at least in the "standard" supernova model, and simply approximating the detector response to $E_\mathrm{vis}=E_\mathrm{\bar\nu_{e}}-0.8$ MeV, one can write: 
\begin{equation}
m_\mathrm{min} = f_\mathrm{bk}\Delta t + M_{\mathrm{act}}~N_\mathrm{p}~ 
\epsilon(E_\mathrm{cut}) 
\int\limits^{10 s}_{0} dt
\int\limits^{100 MeV}_{E_\mathrm{cut}+0.8 MeV}
\Phi(E_{\bar\nu_\mathrm{e}},t) \cdot \sigma(E_{\bar\nu_\mathrm{e}})
dE_{\bar\nu_\mathrm{e}}
\end{equation}
where:
$M_{\mathrm{act}}$ is the active mass, 
$N_\mathrm{p}=9.34~10^{28}$ is the number of free protons in a scintillator ton,
$\epsilon(E_\mathrm{cut})$ is the trigger efficiency approximated as constant ($\epsilon = 0.9$ for $E_\mathrm{cut} = 7$ MeV and $\epsilon = 0.95$ for $E_\mathrm{cut} = 10$ MeV, see fig. \ref{threshold1}),
$\sigma(E_{\bar\nu_\mathrm{e}})$ is the IBD cross section\cite{cross} and
$\Phi(E_{\bar\nu_\mathrm{e}},t)$ the differential $\bar\nu_\mathrm{e}$ intensity at the detector.
The upper limit in the time integral (10 s) corresponds to the maximum unbiased cluster duration.\\
Hence the integral on the right side of (2) is the detector burst sensitivity, $S$,
in terms of minimum neutrino flux times cross section integrated over $\Delta t$ and $\Delta E$, and is expressed as number of neutrino interactions per target:\\
$S_{E_\mathrm{cut}} = (m_{min} - f_\mathrm{bk} \Delta t)/(M_{\mathrm{act}} \cdot N_\mathrm{p} \cdot \epsilon) $\\
The values of $S$ 
are shown in table \ref{table1}, for the two LVD thresholds of the imitation frequency, i.e., $F_{\mathrm{im}}$ = 1 per 100 years and $F_{\mathrm{im}}$ = 1 per month, two different masses, $M_{\mathrm{act}}$ = 1000 t and 330 t, and two values of $E_\mathrm{cut}$.
As it can be seen an important improvement can be obtained by increasing the energy cut from 7 to 10 MeV. As was shown in figure \ref{mult}, the minimum cluster multiplicity, for example at $F_\mathrm{im}$ = 1 per 100 years, goes from 22 to 10, allowing an improvement of almost a factor of two in the sensitivity $S_\mathrm{E_\mathrm{cut}}$.\\
It must be noted that in the on-line algorithm described so far we have neglected the capability of LVD 
to detect both products of the IBD reaction (see section 2.1). We can consider the signature of the reaction to build different burst selection algorithms. For example, we can require that all the H signals in the cluster are "signed", i.e., accompanied by a delayed L one (algorithm IBD-A). 
In the definition of the imitation frequency (eq. 1) f$_\mathrm{bk}$ is substituted by f$_\mathrm{bk}\cdot \bar P_{Lbk}$ (being $\bar P_{Lbk} = 0.13$ the probability for a H signal to be followed, in the same counter, by a L one due to background, averaged over the whole array) and, in (eq. 2), $\epsilon_{E_\mathrm{cut}}$ by $\epsilon_{E_\mathrm{cut}}\epsilon_\mathrm{n}$. 
The sensitivity of the algorithm IBD-A is shown in Table 1: even if the minimum multiplicity is lower, and the background rate is reduced, because of the $n$-capture efficiency ($\epsilon_\mathrm{n}=0.5$) the IBD-A method has comparable effectiveness or even less then the on-line one.\\
We can also build several different algorithms, intermediate between the on-line and the IBD-A ones, requiring that only a fraction of the H signals in the cluster are accompanied by L ones (IBD-B)\footnote{We reject all the clusters which have a number of "signed" H signals $\leq k$, such that: $\sum_{r=0}^{r=k} P(r,m,p) \leq P_0$,
where $P(r,m,p)$ is the binomial probability to have r signed pulses in a cluster of multiplicity $m$. We choose $P_0=0.1$ as an example, the case $P_0=0$ corresponding to the on-line algorithm.}. However, even if the IBD-B method efficiency results higher (as can be seen in Table 1, where an example is given), it does not exceed the on-line one enough to justify the loss of simplicity of the on-line algorithm and its independence from the  model of supernova neutrino emission.
Moreover, the on-line algorithm is sensitive to all possible neutrino interactions in LVD, both in the liquid scintillator and in the iron structure (that can represent up to 15$\%$ of the total number of interactions).\\

\section{Discussion and conclusions}
\begin{figure}[H]
\begin{minipage}{.48\columnwidth}
\includegraphics*[width=0.65\textwidth,angle=90,clip]{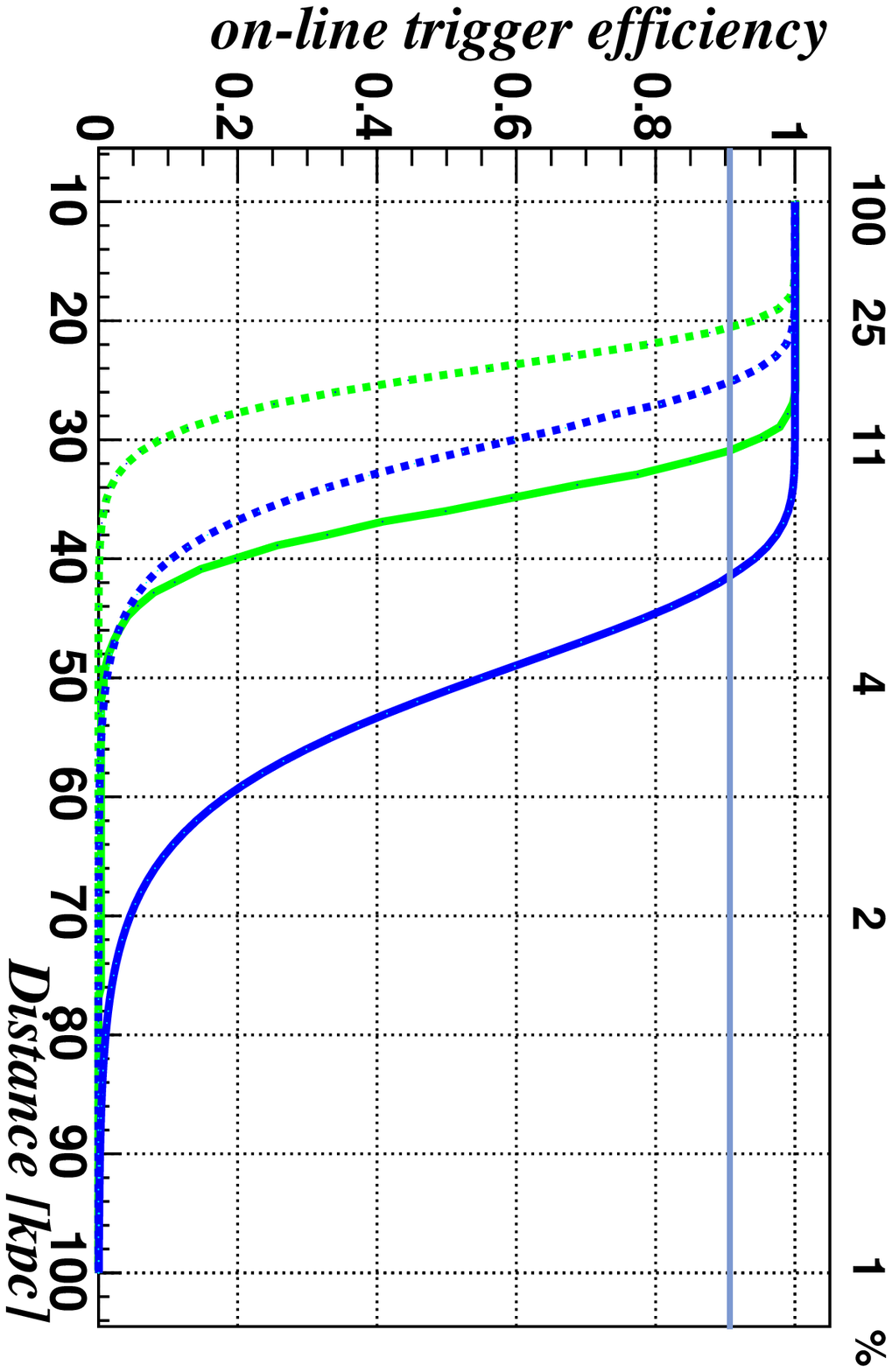}
\end{minipage}
 \hspace{1.5pc} 
\begin{minipage}{.48\columnwidth}
\includegraphics*[width=0.65\textwidth,angle=90,clip]{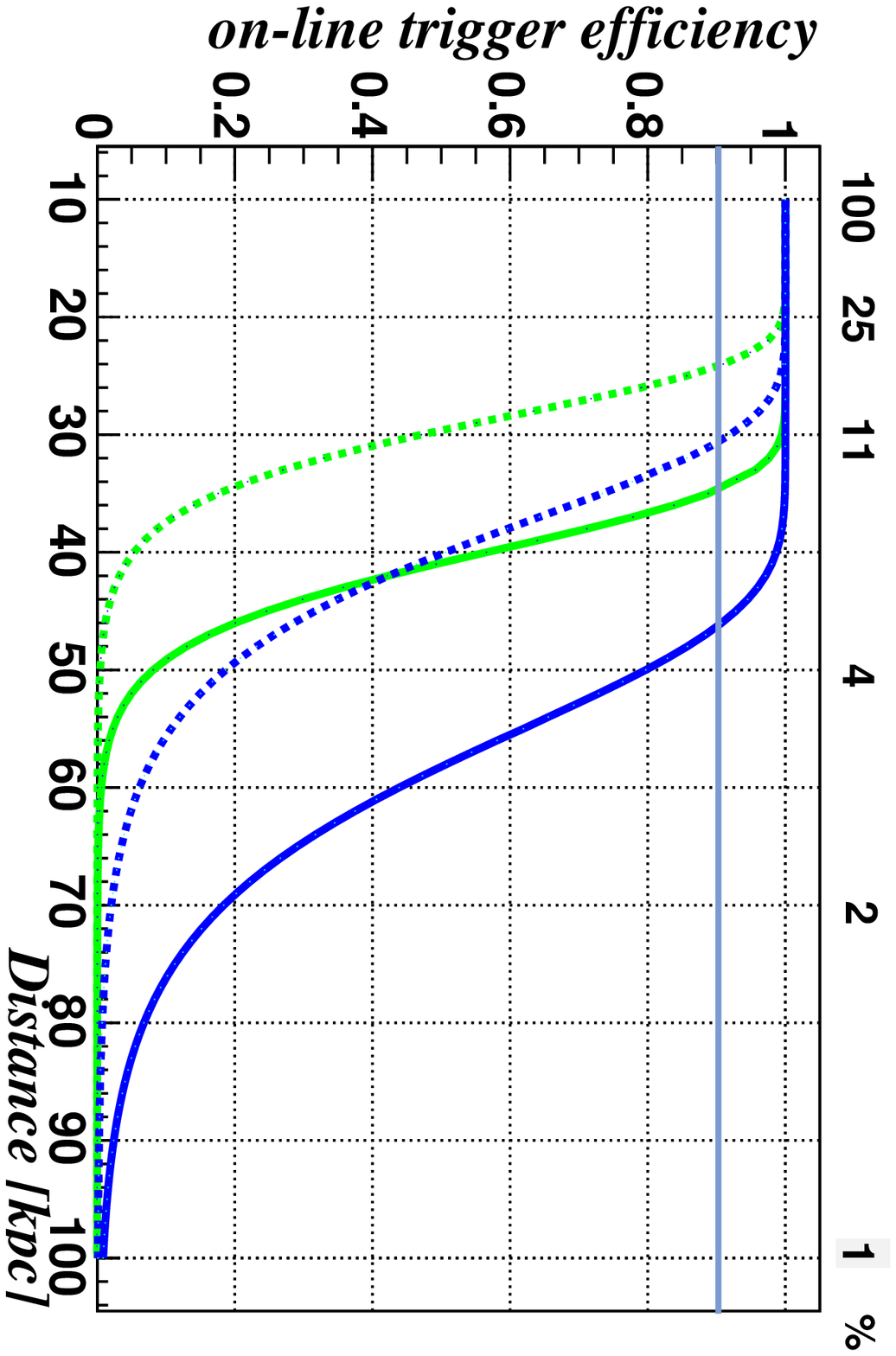}
\end{minipage}
\caption{ {\bf On-line trigger efficiency versus distance (lower scale) and percentage of SN1987A signal at 10 kpc (upper scale) for E$_\mathrm{cut}$=7-10 MeV (light green and dark blue lines, respectively) and M$_\mathrm{act}$ = 330 t (dotted) and 1000 t (continuous). LVD stand-alone on the left panel and in the SNEWS on the right one.}}
\label{eff}
\end{figure}

The Large Volume Detector, located in
INFN Gran Sasso National Laboratory has been designed to detect neutrino bursts from galactic gravitational collapses. In this paper we have described how the trigger and the event selection have been optimized to recognize on-line a neutrino burst from a supernova explosion even in the absence of an "external trigger", such as the presence of an optical counterpart. 
The fast identification of such a neutrino signal is of utmost importance especially in view of the LVD participation to the SNEWS project, which should promptly alert the worldwide network of observatories to allow the study of the rare event since its onset. 
We have discussed the on-line algorithm currently in use at LVD and we have defined its sensitivity S in terms of minimum number of interactions per target unit requested to produce a burst alarm.\\
If one assumes a model for the neutrino emission and propagation, it is possible to express the sensitivity in terms of physical parameters of the source, such as its distance or the emitted neutrino flux.
In particular we will use here the signal detected by Kamiokande-II and IMB for SN1987A.  
Following \cite{Jeger} (see table 1 herein), \cite{Viss} and \cite{Aldo}, we adopt these values for the astrophysical parameters of SN1987A: average $\bar\nu_\mathrm{e}$ energy $<E_{\bar\nu_\mathrm{e}}>$=14 MeV; total radiated energy $E_{b}$ = 2.4$\cdot$10$^{53}$ erg, assuming energy equipartition; distance D=52 kpc and average non-electron neutrino energy 10$\%$ higher than $\bar\nu_\mathrm{e}$\cite{keil}. 
Concerning neutrino oscillations (see \cite{oscill} for a discussion), we consider normal mass hierarchy.
We calculate the number of inverse beta decay signals expected from a SN1987A-like event occurring at different distances, for $E_\mathrm{cut}=7$ and $E_\mathrm{cut}=10$ MeV, and for two values of the detector active mass, M$_\mathrm{act}$ = 330 t and M$_\mathrm{act}$ = 1000 t. 
For example, the number of detected IBD events for such a supernova in the center of the Galaxy (D = 10 kpc) is 230 for E$_\mathrm{cut}$ = 7 MeV and M$_\mathrm{act}$ = 1000 t.
Taking into account Poisson fluctuations in the signal multiplicity, we derive the on-line trigger efficiency as a function of the distance: this is shown in figure \ref{eff} (lower scale) for LVD working stand-alone and in the SNEWS.\\
On the other hand, if we fix the distance of this supernova, e.g. at $10$ kpc, we can derive the LVD sensitivity in terms of the minimum neutrino intensity at the source, varying the total emitted energy. 
The on-line trigger efficiency, as a function of  neutrino luminosity (in terms of percentage of SN1987A one) is shown again in figure \ref{eff}, upper scale.
For example LVD working stand-alone, with 1000 $t$ of active mass and for E$_\mathrm{cut}$ = 10 MeV,  is sensitive (at $90\%$ c.l.) to a neutrino luminosity equivalent to $6 \%$ of that of SN1987A placed at a distance of 10 kpc. 

We can conclude that, without introducing any further check on the time structure, energy spectra and $\nu$ flavor content of the signals in the cluster (which are postponed to the off-line analysis), LVD is able to identify on-line neutrino bursts from gravitational stellar collapses occurring in the whole Galaxy ($D\leq 20$ kpc) with efficiency $>$ 90\%. Such a sensitivity is preserved even if the detector is running with only one third of its total mass and stand-alone, with a severe noise rejection factor ($\leq$1 fake event every 100 years). 
The LVD on-line trigger efficiency can be extended up to 50 kpc (corresponding to the Large Magellanic Cloud) by introducing a cut on the visible energy at 10 MeV.

\section{Acknowledgments}
We thank Francesco Vissani for fruitful discussions.

\newpage

\begin{table}[H]
\caption{Burst sensitivity 
for different selection algorithms, for two energy thresholds E$_\mathrm{cut}$ and for two values of imitation frequency; $M_{act}$ is expressed in ton; 
$m_{min}$ represents the minimum cluster multiplicity (in parenthesis the minimum number of requested "signed" H signals for the IBD-B algorithm).} 
\begin{center}
\begin{tabular}{ c c c c c c c c c c} \hline
&& \multicolumn{3}{c} \textbf{
{$F_{im}=1$ month$^{-1}$} }
 && \multicolumn{3}{c}
  \textbf{{$F_{im}=1\cdot 10^{-2}$ year$^{-1}$}}\\
\hline
E$_\mathrm{cut}$=7 MeV
&
algorithm
&
M$_\mathrm{act}$ 

&
$m_{min}$
&
$S_\mathrm{E_\mathrm{cut}}$
&
&
&
$m_{min}$
& 
$S_\mathrm{E_\mathrm{cut}}$
\\
\hline
\\
&
on-line
&
1000

& $18$ &$1.6\cdot 10^{-31}$ &&& $22$&$2.1\cdot 10^{-31}$&\\
&
&
330

& $10$ &$3.0\cdot 10^{-31}$ &&& $14$&$4.5\cdot 10^{-31}$&\\
\hline\\
&
IBD-A 
& 
1000

& $7$ & $1.5\cdot 10^{-31}$ &&&$10$&$2.2\cdot 10^{-31}$&\\
&
&
330

& $5$ & $3.4\cdot 10^{-31}$ &&&$7$&$4.9\cdot 10^{-31}$&\\
\hline\\
&
IBD-B 
& 
1000 
& $15(5)$ &$1.3\cdot 10^{-31}$&&&$19(7)$&$1.7\cdot 10^{-31}$&\\ 
&
&
330

& $9(3)$ &$2.7\cdot 10^{-31}$&&&$12(5)$&$3.8\cdot 10^{-31}$&\\ 
\hline

\hline
E$_\mathrm{cut}$=10 MeV
&
&
& 
& 
&
&
&
& 
&
\\
\hline
&
on-line
&
1000

& $8$ &$8.3\cdot 10^{-32}$ &&& $10$&$1.1\cdot 10^{-31}$&\\
&
&
330

& $5$ &$1.6\cdot 10^{-31}$ &&& $8$&$2.6\cdot 10^{-31}$&\\
\hline\\

\end{tabular}
\vspace{0.3cm}\\
\end{center}
\label{table1}
\end{table}

\end{document}